\documentclass[aps,pra,preprint]{revtex4-1}
\usepackage[T1]{fontenc}
\usepackage{enumerate}
\usepackage[utf8]{inputenc} 
\usepackage{xcolor}
\usepackage{graphicx}

\newcommand{\bra}[1]{\ensuremath{\left\langle\, #1\,\right|}}
\newcommand{\ket}[1]{\ensuremath{\left|\,#1\,\right\rangle}}

\newcommand{\Tr}[1]{ \mbox{Tr}\left\{ #1 \right\} }

\begin{document}
	\title{Loschmidt Echo and Classicality of the Gamma Model.}
    \author{Gilson V. Soares} 
	\email{gilson.soares@ufrr.br}
	\address{Escola Agrotécnica da Universidade Federal de Roraima , Rodovia
    BR- 174, KM 35, S/N , Murupu, S/N 69310-000 Boa Vista - RR, Brazil }
	\author{Mauricio Reis} 
	\email{mreis@ufsj.edu.br}
	\address{Departamento de F\'isica e Matem\'atica, Universidade Federal de S\~ao
		Jo\~ao Del Rei, C.P. 131, Ouro Branco, MG, 36420 000, Brazil }
	\author{Adelcio C. Oliveira} 
	\email{adelcio@ufsj.edu.br}
	\address{Departamento de F\'isica e Matem\'atica, Universidade Federal de S\~ao
		Jo\~ao Del Rei, C.P. 131, Ouro Branco, MG, 36420 000, Brazil }
	\begin{abstract}
	The classicality of the Gamma Model, an analytically solvable quantum oscillator with non-linear dynamics, is investigated using the overlap dynamics, also known as the Loschmidt Echo, and roughness, a classicality measure based on the Wigner representation of a state. Though the overlap dynamics would indicate a chaotic regime, here the model is integrable. The time mean of the overlap function decays inversely with the effective Hilbert space occupied by the initial state for the non-periodic case. Two different stationary regimes were found for the overlap mean and overlap variance. The state non-classicality was investigated using the roughness measure, and its mean also has two stationary regimes. For large effective Hilbert space, the roughness time mean depends more on the effective space than the initial state. While the Wigner function is dominated by the non-diagonal terms, the overlap operator has some contribution on each part, but it is  more affected by the diagonal therms them the density matrix. 
    
	\end{abstract}
	
	\maketitle

\section{Introduction}

A quantum state under nonlinear dynamics usually presents a collapse of some expectation value, though this is not exactly a quantum behavior but can be regarded as a part of the statistical description \cite{Bal,Ball98,Ball01,Ball2005}. On the other hand, the revival is a quantum feature, and in many cases, its death is used as a classicality criterion \cite{Oliveira09b,adelcio2012,Angelo,Bosco2016,adelcioJMP,Oliveira2021,Kirchmair,Oliveira2014}. For those interested in the problem of the classical limit of quantum mechanics, there are at least two ways to eliminate the revival. The first is through continuous monitoring, and in this case quantum dynamics recovers classical Newtonian trajectories \cite{Oliveira2014,Oliveira2021, Gampel}. The second way is through interaction with an external environment, but the classical limit is now Liouvillian and is only recovered for large classical actions and due to the existence of limitations in experimental resolution \cite{Zur1,adelcio2012,Ball2,Ball3,Ball4,Ball5}.\\

The Loschmidt echo is a measure of the revival of the state when the system is driven in an imperfect reverse dynamics; it is defined by the overlap between two identical initial states evolved by slightly different Hamiltonians \cite{Echo1,Echo2,Echo3}. 
While the  Loschmidt Echo decay is well understood in the chaotic regime, a few things are known for regular systems \cite{Echo2,Echo3}. While the chaotic dynamics have six relevant time
scales, regular dynamics have only three relevant time scales: the classical
averaging time, the quantum fidelity
decay time,  and the stationary regime time \cite{Rev1,Rev2,Prosen}. \\
The overlap decay has several applications; it can be understood as an approximation of the relative entropy between two states, these states being initially equal but evolving according to slightly different dynamics \cite{Vedral,Fanizza,Bartlett}. The probability of survival is an analogous quantity; it is the measure of the probability of the state returning at time t to the initial state \cite{Herrada2023}.
For more details about the overlap, we recommend the instructive reviews about the subject \cite{Rev1,Rev2}.


 Here we investigate a non-linear oscillator $\gamma$-oscillator \cite{Oliveira2021}. It is a non-linear oscillator that has two different regimes: for a positive integer $\gamma$ it is periodic, and for a non-integer, or negative, $\gamma$ its dynamics is complex but the classical counterpart is periodic. As occurs with chaotic models \cite{Echo1}, the overlap for non-integer $\gamma$ dynamics  tends to a small value, with small fluctuations. On the other hand, if $\gamma$ is a positive integer,  the time average of the overlap remains appreciable, and its fluctuations are much larger.\\
In the second part of the work, we define the overlap operator, $\mathcal{K_\rho}$, and investigate in phase space, taking its Wigner transform, how this contributes to the dynamics of the overlap.

\section{The model: Gamma Oscillator}

In this section, we consider the following
quantum Hamiltonian,
\begin{equation}
\hat{H}_{q}=\hbar \omega N+\lambda \hbar ^{2\gamma }\left[ N^{2}+\epsilon %
\right] ^{\gamma },
\end{equation}
where $\omega$ is the harmonic frequency, $\lambda$ is the non-linear 
coupling constant, $\gamma$ and $\epsilon$ to  are free parameters, with the 
restriction $\epsilon > 0$ and $N$ is the usual number operator of the harmonic oscillator.
For this model, neither large action nor complex dynamics due to nonlinearities are relevant to the process when trying to recover the Classical Limit of Quantum Mechanics \cite{Oliveira2021}.
For $\gamma =1$ the model of the quartic oscillator,  see \cite{Ber77,Ber81,Adelcio2003,Oliveira06,Oliveira09b,adelcio2012,Faria,Ber91,Leonski1996,Iomin2001,Iomin2003}. The collapse and revival time scales depend on the initial state, and the parameter and
its typical time scales can vary several 
orders of magnitude \cite{Bosco2016}.

Considering an initial state given by%
\begin{equation}
\rho_0 =\sum_{m,n=0}^{\infty} A_{m,n}\left\vert m\right\rangle\bra{n} ,
\end{equation}%
where $\ket{k}$ is the number operator ($N$) eigenstate. It's time evolution is
\begin{equation}
\rho_ t =\sum_{m,n=0}^{\infty }e^{-\frac{it}{\hbar }%
\left[ \hbar \omega (m-n)+\lambda \hbar ^{2\gamma }\left[\left( m^{2}+\epsilon \right)
^{\gamma }-\left( n^{2}+\epsilon \right)
^{\gamma }\right)\right] }A_{m,n}\left\vert m\right\rangle\bra{n} .
\end{equation}%

\section{Loschmidt Echo}

To analyze the overlap dynamics, we initially set two different Hamiltonians 
differing only by the $\lambda$ value, i.e. $H_1=H(\lambda_1)$ 
and $H_2=H(\lambda_2)$, will be considered. 
A Hermitian operator of particular interest is the difference between them:
\begin{equation}
\Delta H=(\lambda_1-\lambda_2) \hbar ^{2\gamma }\left[ N^{2}+\epsilon %
\right] ^{\gamma },
\end{equation}
where each constant has the same meaning explained before. The term $(\lambda_1-\lambda_2)  \hbar ^{2\gamma }$ can be viewed as a time scale factor and, 
for practical reasons, was set to $1$ in the numerical evaluations.
The overlap, or the Fidelity, between the two different time evolutions of the same initial state can be calculated by
\begin{equation}
    O_{\Delta H}=\left(\Tr{\sqrt{\sqrt{\rho_1(t)} \rho_2(t)\sqrt{\rho_1(t)}}}\right)^2.
\end{equation}
For an initial pure state $\ket{\psi_0}$, then the Loschmidt echo ($O_{\Delta H}$) is given by 
\begin{equation}
     O_{\Delta H}(\psi_0)=|\bra{\psi_0}e^{i\frac{H_2t}{\hbar}}e^{-i\frac{H_1t}{\hbar}}\ket{\psi_0}|^2
\end{equation}
as it was originally defined \cite{Echo1,Echo2,Echo3}.
Here, $[H_1,H_2]=0$ and $\rho_1(0)=\rho_2(0)=\rho(0)=\ket{\psi_0}\bra{\psi_0}$, the overlap is readily obtained using the interaction picture:
\begin{equation}
    O_{\Delta H}=\Tr{\rho(0) \rho_{\Delta}(t)}, 
    \label{LoschmidfEco}
\end{equation}
where $\rho_{\Delta}(t)=U_{\Delta H}\rho(0)U^{\dagger}_{\Delta H }$,  $U_{\Delta H}=exp\left[\frac{-it}{\hbar}\Delta H\right]$.

The Loschmidt echo quantifies the degree of irreversibility, while the fidelity is a measure of the sensitivity of quantum evolution to perturbations. For pure states,   the Loschmidt echo and fidelity are equivalent \cite{Echo3}. We use the definition of \ref{LoschmidfEco}, and consequently it is the Loschimid echo, but it can be interpreted as fidelity for pure states.

\begin{figure}[!htbp]
\centering
\includegraphics[width=14cm,height=10cm]{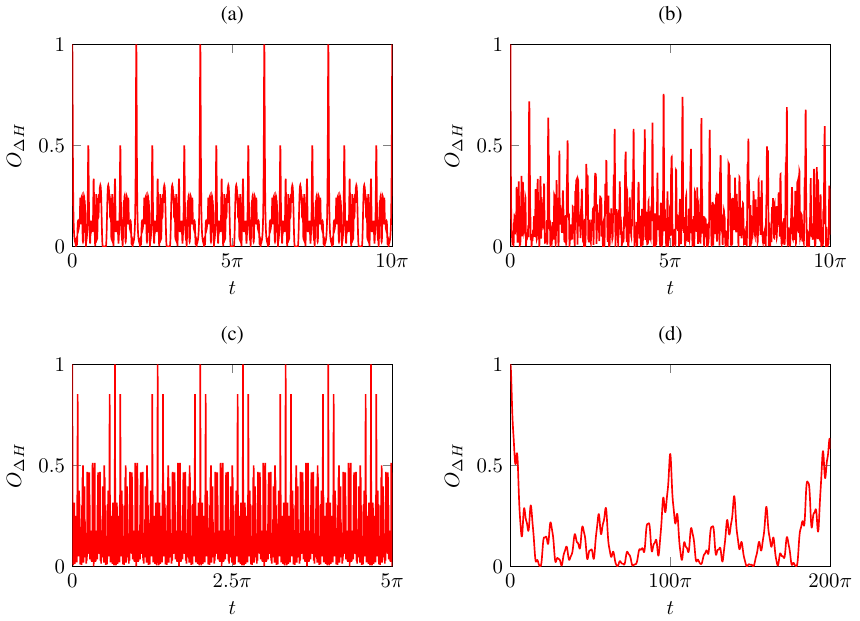}
\caption{Overlap function $ O_{\Delta H}$ as function of time for $\epsilon=1$, $\hbar=1$ and for a coherent state, ($\ket{\alpha}$) as initial state with $\alpha =2$. (A) $\gamma=1$, (B) $\gamma=1.1$, (C) $\gamma =2$ and (D) $\gamma=-1$.}
\label{Ov1}
\end{figure}

The overlap dynamics ($ O_{\Delta H}$) is periodic if $\gamma$ is a positive integer; otherwise, it has many incommensurable frequencies, and the complete revival is absent.  Figure \ref{Ov1} shows the $ O_{\Delta H}$ as a function of time for (A) $\gamma=1$, (B) $\gamma=1.1$, (C) $\gamma =2$ and (D) $\gamma=-1$. (A) and (C) are periodic, and (B) and (D) aren't. The initial state is a coherent state with $\alpha=2$. 

\begin{figure}[!htbp]
\centering
\includegraphics[width=14cm,height=10cm]{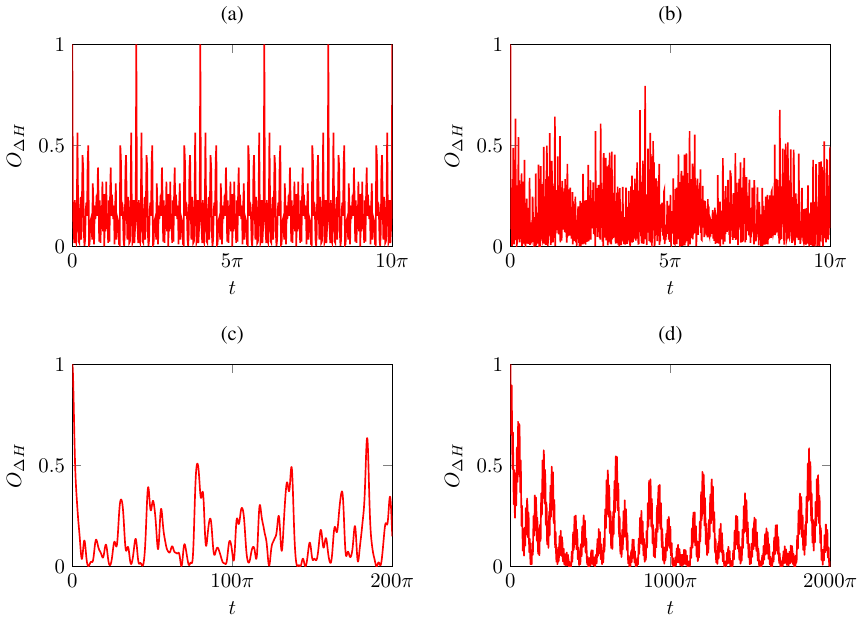}
\caption{Overlap function $ O_{\Delta H}$ as function of time for $\epsilon=1$, $\hbar=1$ and for a coherent state ($\ket{\alpha}$) as initial state with $\alpha =2$. (A) $\gamma=3$, (B) $\gamma=3.1$, (C) $\gamma =-0.5$ and (D) $\gamma=-2$.}
\label{Ov2}

\end{figure}

In figure \ref{Ov2} the overlap, $ O_{\Delta H}$, is presented for (A) $\gamma=3$, (B) $\gamma=3.1$, (C) $\gamma =-0.5$ and (D) $\gamma=-2$, as  \ref{Ov2}-A shows, there is a revival for $t=2 \pi$ and the dynamics is periodic, as the $|\gamma|$ increases the overlap periodicity is less evident. 

It is known that there is a significant difference in the time mean of the overlap due to the classical analog regime being smaller when it is chaotic, and also its variance \cite{Echo3,Hilb1,Hilb2,Hilb3,Hilb4}. A question that can be done is, are the chaotic dynamics responsible to smaller fluctuations in the overlap, or is it due to the fact that the overlap spectra are not commensurable? As the classical counterpart of our model has no chaos, we can only conjecture that the quantum dynamics should have three different regimes: the chaotic, periodic, and non-periodic.

\begin{figure}[!htbp]
\centering
\includegraphics[width=14cm,height=6cm]{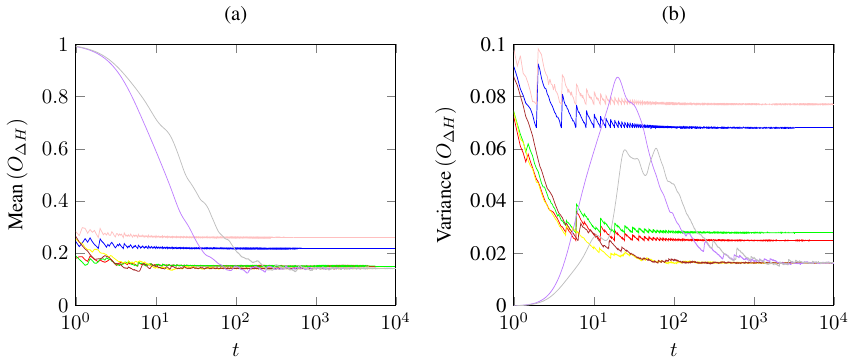}
\caption{Time mean of the overlap function $ O_{\Delta H}$ as a function of time for $\epsilon=1$, $\hbar=1$ and for a coherent state ($\ket{\alpha}$) as the initial state with $\alpha =2$. In red $\gamma=1$, blue $\gamma=2$, green $\gamma =3$, pink $\gamma=4$, yellow $\gamma=1.7$, brown $\gamma=3.5$, purple $\gamma=-0.5$ and gray $\gamma=-1$.  }
\label{Ov3}
\end{figure}


In figure \ref{Ov3}-a a time mean of the overlap function $ O_{\Delta H}$ as a function of time is presented for many values of $\gamma$. The short-time behavior is a decaying regime $\propto e^{-\Gamma t}$ for non-integers $\gamma$. We can identify two asymptotical regimes: for negative or non-integer values, $\gamma$ the overlap mean converges to a value around $0.14$, and for $\gamma$ positive integers, it converges at larger values, and apparently it does not have a regular pattern. It is possible to conclude that in the periodic regime, the time mean   $ O_{\Delta H}$ saturates at a higher value, and thus, it can be considered as a witness of periodicity.

Figure \ref{Ov3}-b shows the time variance of the overlap function $ O_{\Delta H}$ as a function of time for some values of $\gamma$. As occurs with the mean (figure \ref{Ov3}), the variance decays exponentially in the short time regime and saturates for long times. Again, for positive integer $\gamma$ it saturates at higher values, and the non-periodic regime saturates at a smaller value.
Also we observe that negative values of $\gamma$ demands a bigger saturation time. 
\begin{figure}[!htbp]
\centering
\includegraphics[width=14cm,height=6cm]{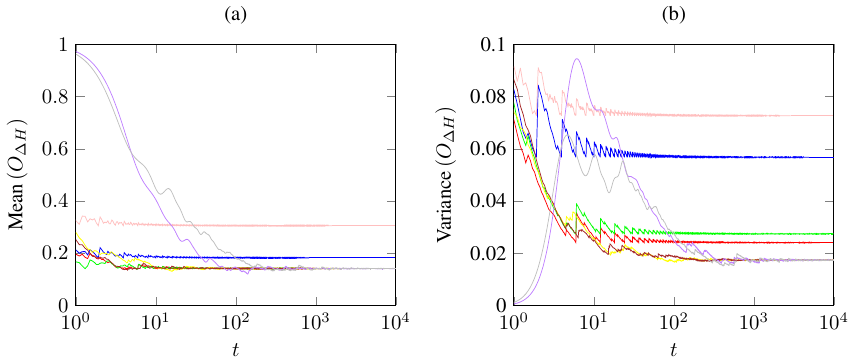}
\caption{(A) - Time mean of the overlap function $ O_{\Delta H}$ as a function of time. (B) Time variance of the overlap function $ O_{\Delta H}$ as a function of time. The initial state is a phase state $\ket{\phi}=\frac{1}{\sqrt{r+1}}\sum_{k=0}^r \ket{k}$, for $r=6$.In red $\gamma=1$, blue $\gamma=2$, green $\gamma =3$, pink $\gamma=4$, yellow $\gamma=1.7$, brown $\gamma=3.5$, purple $\gamma=-0.5$ and gray $\gamma=-1$. }
\label{Ov6}
\end{figure}


The overlap behavior is usually affected by the initial state \cite{Rev1,Rev2}. To observe the effect of the initial state in the overlap dynamics, we use the Pegg-Barnett state \cite{Pegg} known as the phase state. While the coherent state is the least quantum among the pure states, this state is among those with the highest non-classicality \cite{NCMRev}. The phase state was proposed in  order to study the problem of definition and measurement of the
phase of the quantum electromagnetic field \cite{Dirac}.  In figure \ref{Ov6},  the time mean (A) and time variance (B) of the overlap function $ O_{\Delta H}$ are plotted as a function of time for a phase state, the initial state is a phase state $\ket{\phi}=\frac{1}{\sqrt{r+1}}\sum_{k=0}^r \ket{k}$, for $r=6$.  The asymptotic regime depends on $\gamma$, for non-integer values of $\gamma$ the behavior is analogous of the coherent state, while for integer values of $\gamma$ is periodic and behavior is sensitive to the initial state.

\begin{figure}[!htbp]
\centering
\includegraphics[width=12cm,height=6cm]{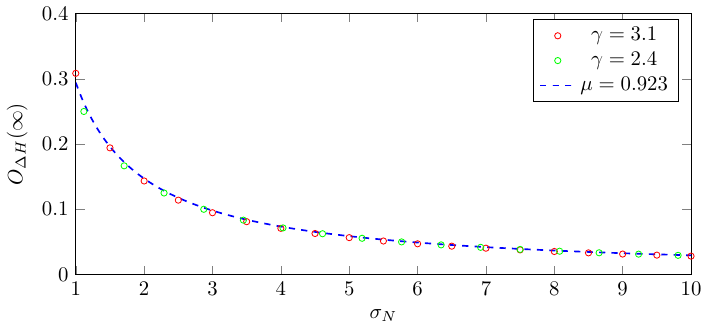}
\caption{Time mean of the overlap function $ O_{\Delta H}$ for $t \rightarrow \infty$ as a function of deviation of the operator $N$ in the state $\rho(0)$ time for $\epsilon=1$.  Red circles $\gamma=3.1$ and green $\gamma=2.4$. The blue dotted line is the fitting function $Z$ equation \ref{Zf}  for $\mu =0.923$. The initial state is a phase state for the circles and a coherent state for the cross.}
\label{Ov7}
\end{figure}

The overlap dynamics, at long time regime and for non-integer $\gamma$, presents a saturation value that is correlated with the effective occupied Hilbert space size, as occurs with full chaotic dynamics \cite{Echo3}. The effective Hilbert space (HS) is given by
\begin{equation}
    Hs=\sqrt{1+12\sigma_N^2},
\end{equation}
where $\sigma_N^2=Tr(\rho N^2)-Tr(\rho N)^2$, and $N$ is the number operator; see reference \cite{Oliveira09b} for details. In figure \ref{Ov7}, the time mean of the overlap function $ O_{\Delta H}$ for $t \rightarrow \infty$ as function of deviation o of the operator $N$, the saturation value can be fitted by $Z$, and it is given by
\begin{equation}
    Z(\sigma)=\frac{\mu}{\pi \sigma_N}.
    \label{Zf}
\end{equation}

The long-time regime is shown for an initial phase state and for coherent state; for non-integers, $\gamma$ the saturation value is correlated with the effective occupied Hilbert space size. Note that this result is independent of the initial state and not sensitive to the non-linearity, $\gamma$, was observed.
Even though the system dynamics are not chaotic, there is an apparently chaotic behavior in the overlap dynamics.

In table \ref{table:alpha} the convergence values  of $O_{\Delta H}(\infty)$, and $O_{\Delta H}(\infty)$ for different values of $\gamma$ and for a coherent state $\alpha=2$, that corresponds to an effective Hilbert space of $H_s=7$. Table \ref{table:r} shows the equivalent for an initial phase state with the same $H_s$. As we can see, the same pattern is observed in both tables; integer values of $\gamma$ tend to converge to distinct values, both for the mean of the overlap and its variance. Although the values are slightly different, for non-integer and negative $\gamma$ values, there is convergence in the asymptotic values. The time variance of the overlap is more sensitive to the exact form of the initial state than its time mean.

\begin{table}
    \centering
    \begin{tabular}{|c|c|c|c|}
    \hline
         $\gamma$   & $\mbox{Mean of}\, O_{\Delta H}(\infty)$ &$\mbox{Variance of}\, O_{\Delta H}(\infty)$  \\
         \hline
        1 &0.1524  &0.0249 \\
        2 &0.2189  &0.0678 \\
        3 &0.1524  &0.0278  \\
        4 &0.2607  &0.0768  \\
        1.7 &0.1434  &0.0164 \\
        3.5 &0.1434  &0.0164 \\
        -0.5 &0.1434  &0.0164 \\
         -1 &0.1440  &0.0166 \\
        \hline
    \end{tabular}
    \caption{Values of convergence of the time mean of $O_{\Delta H}(\infty)$, and variance of   $ O_{\Delta H}(\infty)$ for an initial coherent state  $\alpha =2$ that corresponds to $H_s=7$.}
    \label{table:alpha}
\end{table}

\begin{table}
    \centering
    \begin{tabular}{|c|c|c|}
    \hline
         $\gamma$   & $\mbox{Mean of}\, O_{\Delta H}(\infty)$ &$\mbox{Variance of}\, O_{\Delta H}(\infty)$  \\
        \hline
        1 &0.1428  &0.0241  \\
        2 &0.1836  &0.0566 \\
        3 &0.1428  &0.0274 \\
        4 &0.3061  &0.0725 \\
        1.7 &0.1428  &0.0174 \\
        3.5 &0.1428  &0.0175 \\
        -0.5 &0.1429  &0.0175 \\
        -1 &0.1432  &0.0175 \\
        \hline
    \end{tabular}
    \caption{Values of convergence of the time mean of $O_{\Delta H}(\infty)$, and  variance of $ O_{\Delta H}(\infty)$ for an initial phase state with $r =6$ that corresponds to $H_s=7$.}
    \label{table:r}
\end{table}

\section{Loschmidt echo decay and classicality}
The behavior of the overlap function is similar to that of the classical counterpart for noninteger $\gamma$. The classical Liouvillian dynamics, for non-harmonic systems, causes the classical probability density to spread throughout the accessible phase space, meaning there is no revival. Consequently, the revival is a quantum signature \cite{Oliveira06}.
The classical behavior of the overlap is characterized as a consequence of the dynamics and does not depend on the initial state. In this case, the overlap decay is independent of the $\gamma$ value for nonintegers $\gamma$, but is it really a manifestation of a classical behavior of the state?  To answer this question, let us introduce the Roughness \cite{Lemos2018,Karen2}, it is a distance measure between Wigner Function (W) and Husimi function (H), for a state, $\rho$ it is defined as
\begin{equation}
    R(\rho)^2=2\pi \int dqdp |W_\rho(q,p)-H_\rho(q,p)|^2
    \label{Rug}.
\end{equation}

The Roughness has some remarkable properties; let's summarize them as
\begin{enumerate}
    \item $R \in [0,1]$;
    \item $R(\rho)$ is limited by the purity, i.e., $R(\rho)^2\le \Tr{(\rho ^2)};$
    \item  $R \rightarrow 0$ for thermal states as $T\rightarrow \infty$;
    \item $R \rightarrow 1$ for squeezed states (in the infinity squeezing parameter) and fock states as the principal quantum number $n\rightarrow \infty$;
    \item The coherent state has the minimal value of Roughness between pure states, $R_c=\frac{1}{\sqrt{6}}$;
    \item The superposition states $\ket{\alpha} \pm \ket{-\alpha}$ saturate below unity, $\lim_{\alpha \rightarrow \infty}R=\sqrt{\frac{7}{12}}$;
    \item For a qubit in the Fock state basis, there is a complementary relation between roughness, concurrence (or linear entropy for nonpure states), and Mandel's Q parameter \cite{Reis2018,Reis2021,Reis2021b};
    \item For any incoherent mixture  $\sigma_\lambda = \lambda \rho_1 + (1 - \lambda) \rho_2$ satisfies $\frac{d^2 R({\sigma_\lambda})}{d\lambda^2} \ge 0$;
    \item For pure states, without squeezing, roughness and Wigner negativity are strongly correlated \cite{NCMRev}.
    
\end{enumerate}

For a numerical implementation code of roughness, see \cite{ReisQutip}, and a detailed comparison of the most common non-classical measures can be found in \cite{NCMRev}. 



\begin{figure}[h]
\centering
\includegraphics[width=14cm,height=5cm]{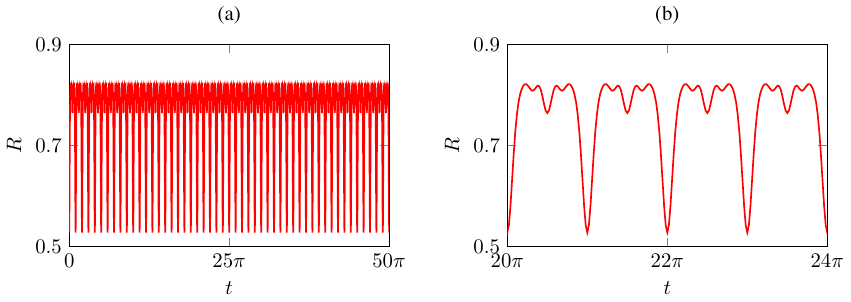}
\caption{(a) Roughness as a function of time ($t$) a for the initial phase state  $\ket{\psi}=\sum_{n=0}^{5}\ket{n}$, for $\gamma=1$, $\lambda=1$, $\omega=0$ and $\epsilon=1$. (b) is the same as (a) for a different time interval.}
\label{Rfase1}

\end{figure}

\begin{figure}[h]
\centering
\includegraphics[width=14cm,height=5cm]{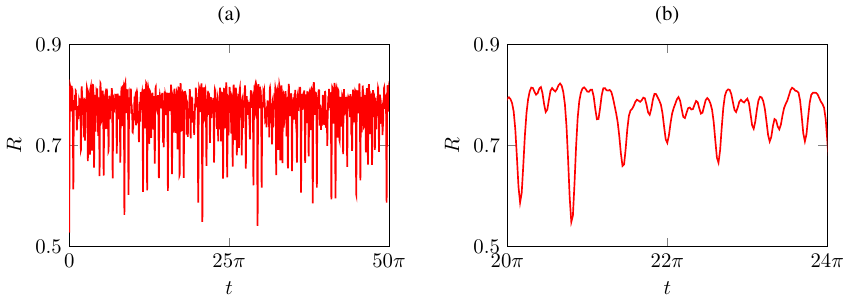}
\caption{(a) Roughness as function of time ($t$) for the initial phase state  $\ket{\psi}=\sum_{n=0}^{5}\ket{n}$, for $\gamma=1.1$, $\lambda=1$, $\omega=0$ and $\epsilon=1$. (b) A part of the time interval of (a).}
\label{Rfase2}

\end{figure}

Considering that cat states have a roughness limited to $\sqrt{\frac{7}{12}}\approx 0.76376$, then a $R>0.7$ can be considered as strongly non-classical. In figure \ref{Rfase1} it the roughness is shown as a function of time ($t$) for the initial phase state, and $\gamma=1$, as can be observed, the roughness achieves values greater than $0.7$ in a large time interval; this is the periodic case. For this figure and the following, we used $\omega=0$, as the non-harmonic term commutes with the harmonic one, then we can consider a rotating frame of angular speed $\omega$, and the plots become less ``noisy.'' A similar behavior of \ref{Rfase1} can be observed in Figure \ref{Rfase2} for $\gamma=1.1$ the non-periodic case; thus, the final state cannot be considered as classical. In fact, this apparent classical behavior is one case where a specific observable can have an apparently classical behavior, but it does not mean that the state is classical \cite{Adelcio2003}.

\begin{figure}[h]
\centering
\includegraphics[width=14cm,height=6cm]{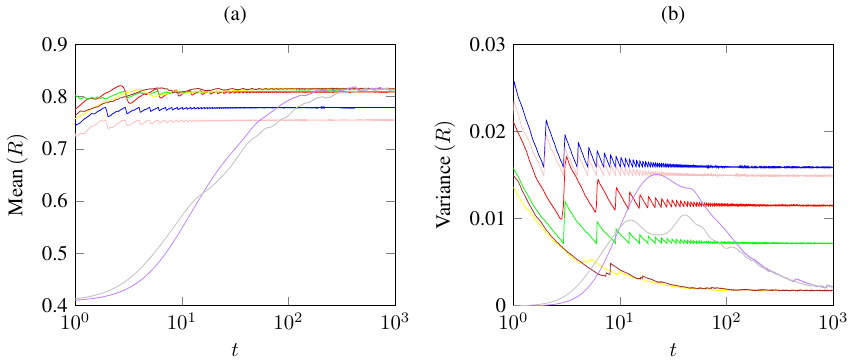}
\caption{(a) Time mean roughness as a function of time ($t$) for the initial coherent state  $\alpha=2$, for $\gamma=1_1$, $\lambda=1$, $\omega=0$ and $\epsilon=1$. (b) Same as (a) for longer times.In red $\gamma=1$, blue $\gamma=2$, green $\gamma =3$, pink $\gamma=4$, yellow $\gamma=1.7$, brown $\gamma=3.5$, purple $\gamma=-0.5$ and gray $\gamma=-1$. }
\label{Rfase3}
\end{figure}

In figure \ref{Rfase3}-(a) the time mean roughness, $\overline{R}$, is shown, and in (b) its time variance $\Delta(R)$.  For all $\gamma$ values $\overline{R}$ converges for a value greater than $0.7$ with small $\Delta(R)$. For non-integer $\gamma$, the values of $\Delta(R)$ converge to the same value, indicating the same quantumness regime.
For each integer value of $\gamma$ it  converges at different values. In fact, we observed five asymptotical regimes, non-integer and negative values of $\gamma$, and four values of convergence for positive integers. In our numerical simulations, the positive integer falls ``approximately'' on one of the four curves.

\begin{figure}[h]
\centering
\includegraphics[width=14cm,height=6cm]{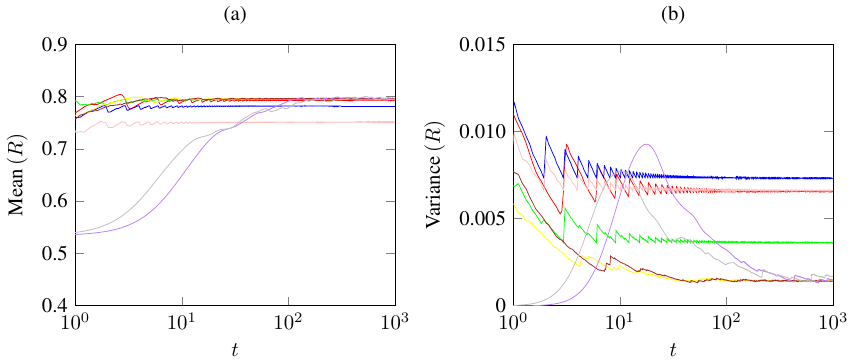}
\caption{(a) Time mean roughness as a function of time ($t$) for the initial phase state  $\ket{\psi}=\sum_{n=0}^{6}\ket{n}$, for $\gamma=1_1$, $\lambda=1$, $\omega=0$ and $\epsilon=1$. (b) Same as (a) for longer times. In red $\gamma=1$, blue $\gamma=2$, green $\gamma =3$, pink $\gamma=4$, yellow $\gamma=1.7$, brown $\gamma=3.5$, purple $\gamma=-0.5$ and gray $\gamma=-1$. }
\label{Rfase4}
\end{figure}

In figure \ref{Rfase4} we show the same graph of \ref{Rfase3} for an initial phase state. Although the figures are not exactly equal, they are qualitatively equivalent. Here we also have five regimes with the same properties.

\begin{figure}[h]
\centering
\includegraphics[width=12cm,height=6cm]{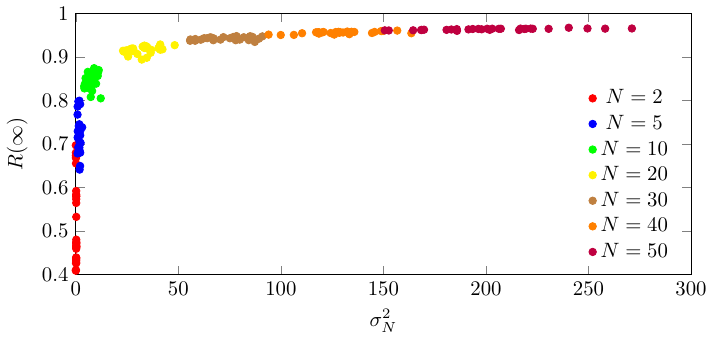}
\caption{Time mean of roughness for $t\rightarrow \infty$, for $\gamma=1.7$. $N$ is the basis size used.}
\label{fig:Ruginfty}
\end{figure}

In figure \ref{fig:Ruginfty} we show the time mean of roughness  for $t\rightarrow \infty$, for $\gamma=1.7$. $N$ is the basis size used. For each value of $N$ we generate $50$ random states and evolve it to a large time. For a large effective Hilbert space, or $\sigma_N$, we see that the mean grows monotonically and its values do not depend significantly on the initial state.

\section{Quantum structures in phase space}
As was clearly shown in the previous section, the overlap decay does not imply classicality; in fact, the non-classical indicator increases in the non-periodic regime. This is a consequence of a well-known quantum behavior when the classical limit is investigated, which is the fact that a specific observable behaving classically does not necessarily imply that the quantum state is classical \cite{Adelcio2003,Oliveira06,adelcio2012,Lemos2018}. Due to the overlap definition, equation \ref{Ov3}-b, its value depends strongly on the diagonal terms of the density matrix. In practical terms, it means that the overlap does not change if a dispersive environment (or a phase reservoir) is included \cite{Oliveira09b}. In order to understand the implications of suppressing the non-diagonal density matrix, let us analyze the Wigner function. For quantum states defined in the Fock basis, it is straightforward to obtain the Wigner function \cite{Faria}, it is given by
\begin{equation}
W(\rho ,\epsilon ,\gamma )=\frac{\sum {}_{n,m=0}^{\infty }A_{n,m}e^{it\left[
\omega (m-n)+\lambda \hbar ^{2\gamma -1}\left( m^{2}+\epsilon \right)
^{\gamma }-\lambda \hbar ^{2\gamma -1}\left( n^{2}+\epsilon \right) ^{\gamma
}\right] }\Pi _{n,m}}{2\pi },
\end{equation}%
where  $\Pi _{n,m}$ is given by 

\begin{eqnarray}
\Pi _{m,n}\left( \beta \right) & =\frac{\left( -1\right) ^{m}}{\pi \hbar }%
\sqrt{\frac{2^{n-m}m!}{n!}}e^{-\left\vert \beta \right\vert ^{2}}\beta
^{n-m}L_{m}^{n-m}\left( 2\left\vert \beta \right\vert ^{2}\right) ;:n\geq m,
\label{Pi_mn2a} \\
\Pi _{m,n}\left( \beta \right) & =\frac{\left( -1\right) ^{n}}{\pi \hbar }%
\sqrt{\frac{2^{m-n}n!}{m!}}e^{-\left\vert \beta \right\vert ^{2}}\left(
\beta ^{\ast }\right) ^{m-n}L_{n}^{m-n}\left( 2\left\vert \beta \right\vert
^{2}\right) ;:m>n. 
\label{Pi_mn2}
\end{eqnarray}

The functions $L_{m}^{n-m}\left( x\right) $ are the associated Laguerre polynomials.

\begin{figure}[h]
\centering
\includegraphics[width=15cm,height=10cm]{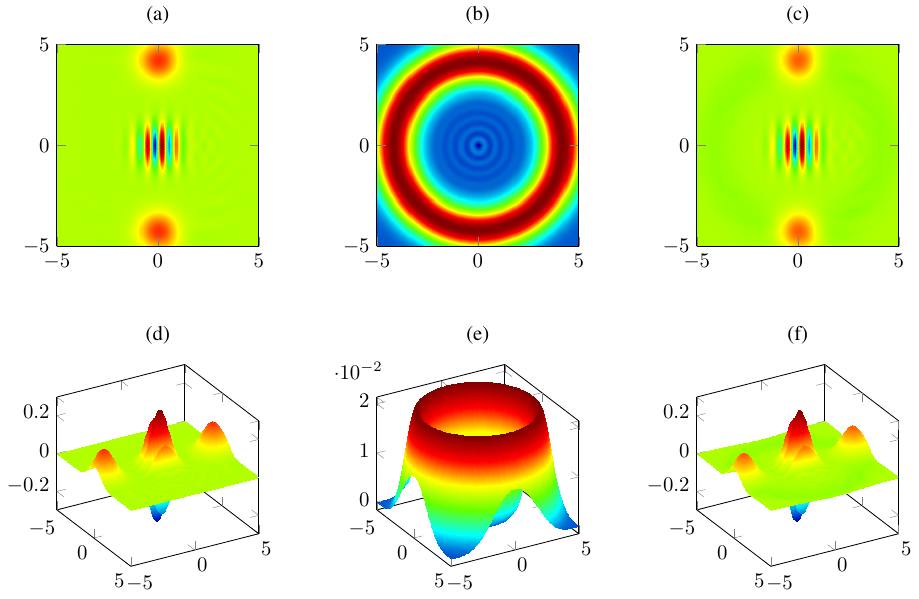}
\caption{ Wigner function for $t=\pi/2$, a coherent initial state with $\alpha=3$, $\epsilon=1$, $\gamma=2$ and $\hbar=1$. In (a) is the contour plot, and (d) is the surface plot of the Wigner function. In-(b) is shown the contour plot, and (e) is the surface plot of diagonal terms of the Wigner function. In (c) is the contour plot, and (f) is the surface plot of non-diagonal terms of the Wigner function.}
\label{WIG1}
\end{figure}

Now let's consider the density matrix \ref{Ov3} written as
\begin{equation}
    \rho_t=\rho_t^D+\rho_t^{ND},
\end{equation}

where $\rho_t^D=\sum_n A_{n,n}(t)\ket{n}\bra{n}$ and $\rho_t^{ND}=\sum_{n \neq m} A_{n,m}(t)\ket{n}\bra{m}$, correspondingly to the diagonal terms of $\rho$ and non-diagonal terms of, $\rho$ respectively. The Wigner function can also be defined by a sum of two contributions $W_D(t)$ and $W_{ND}(t)$ representing the corresponding Wigner term of the diagonal part and non-diagonal part, respectively.
Figure \ref{WIG1} shows the Wigner function for $t=\pi / 2$, $\alpha=3$, $\epsilon=1$ and $\gamma=2$. In figure \ref{WIG1}-(a) is the contour plot, and \ref{WIG1}-(d) is the surface plot of the Wigner function. The \ref{WIG1}-(b) shows the contour plot, and \ref{WIG1}-(d) is the surface plot of diagonal terms of Wigner function, while \ref{WIG1}-(c) is the contour plot, and \ref{WIG1}-(f) is the surface plot of non-diagonal terms of the Wigner function.

\begin{figure}[h]
\centering
\includegraphics[width=15cm,height=10cm]{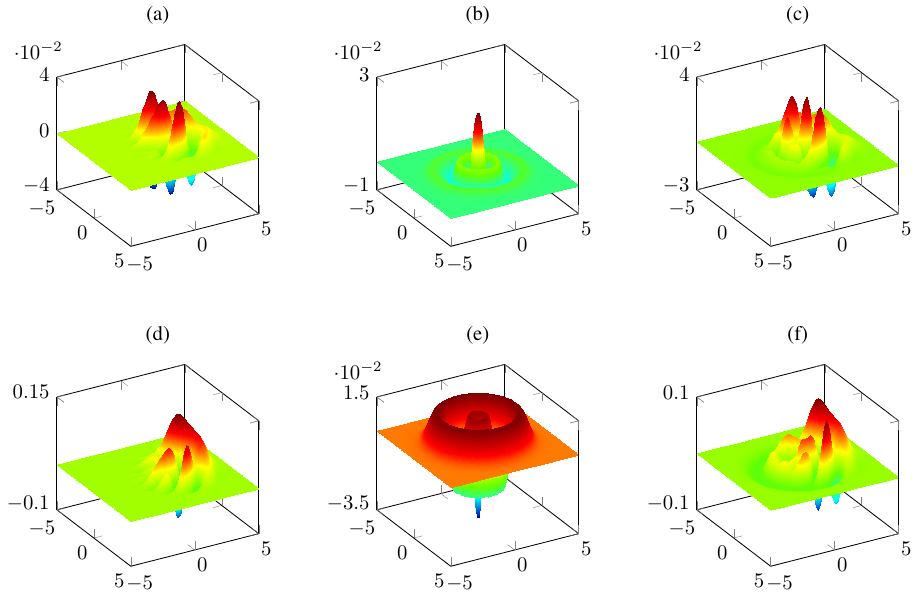}
\caption{ Wigner transform of $\mathcal{R}_\rho$  for $t=5\pi$,a coherent initial state with $\alpha=2$, $\epsilon=1$, $\gamma=1.7$ and $\hbar=1$. In (a) and (d) is the surface plot of the total Wigner transform of  $\mathcal{R}_\rho$. In (b) and (e)  is the surface plot of the Wigner transform of diagonal terms of  $\mathcal{R}_\rho$. In (c) and (f) is the surface plot of the Wigner transform of non-diagonal terms of  $\mathcal{R}_\rho$. (a), (b) and (c) for $t=5\pi$ and (d), (e) and (f) for $t=10 \pi$.
}
\label{WIG2}
\end{figure}

\begin{figure}[h]
\centering
\includegraphics[width=15cm,height=10cm]{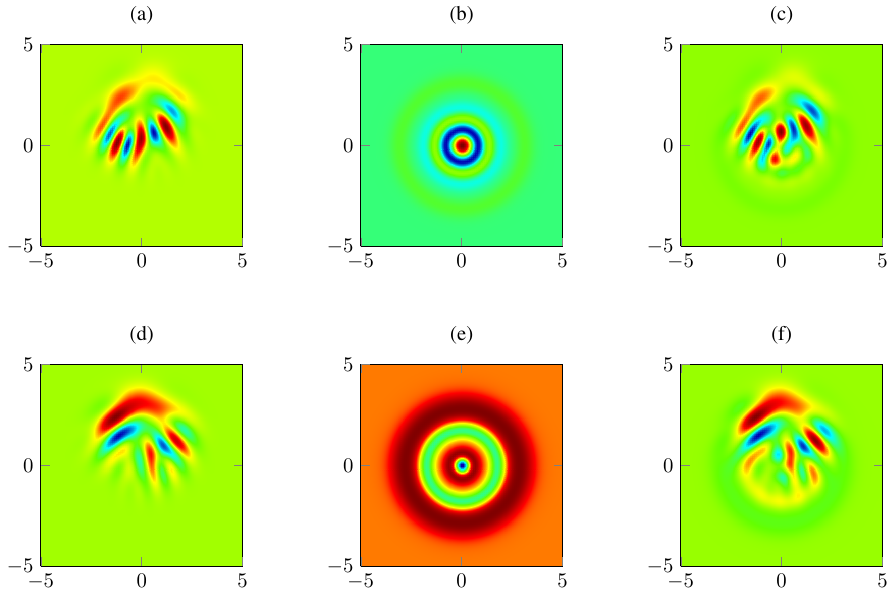}
\caption{ Wigner transform of $\mathcal{R}_\rho$  for $t=\pi$,a coherent initial state with $\alpha=2$, $\epsilon=1$, $\gamma=1.7$ and $\hbar=1$. In (a) and (d) is the contour plot of the total Wigner transform of  $\mathcal{R}_\rho$. In (b) and (e)  is the contour plot of the Wigner transform of diagonal terms of  $\mathcal{R}_\rho$. In (c) and (f) is the contour plot of the Wigner transform of non-diagonal terms of  $\mathcal{R}_\rho$. (a), (b) and (c) for $t=5\pi$ and (d), (e) and (f) for $t=10 \pi$.
}
\label{WIG3}
\end{figure}

The non-diagonal terms of the Wigner function are responsible for the negativity and for the sub-Planck structure. The inner blue circle in figure \ref{WIG1}-(b) has a size of the order of a coherent state, and then the small structures that appear in \ref{WIG1}-(a) and \ref{WIG1}-(c) are contentedly sub-Planck structures. As the overlap is less sensitive to the non-diagonal terms, then it essentially depends more on ``classical'' content of the state. In fact, the overlap is not a function of $\rho(t)$, instead it is a function of $\rho(0)\rho_{\Delta}(t)$ as given by equation \ref{Ov5}. Then let's define an operator $\mathcal{R}_\rho$ as
\begin{equation}
    \mathcal{R}_\rho(t)=\frac{[\rho(0)\rho(t)+\rho(t)\rho(0)]}{\mathcal{K}} ,
\end{equation}
where $\mathcal{K}$ is defined by $tr\{\mathcal{R_\rho}(0)\}=1$. For pure staes, the overlap can be obtained directly by $O_{\Delta H}=tr\{\mathcal{R}_\rho\}$. 
The ``Eigenstate Thermalization Hypothesis'' (ETH)  is defined for an isolated quantum system of many particles; for some classes of initial states, the expectation value will approach thermal equilibrium. It is a universal phenomenon and can be understood as a memory loss of the initial state \cite{Thermal6}, also the long-time saturation occurs at a value inversely proportional to the effective size of the Hilbert space of the system \cite{Echo3,Hilb1,Hilb2,Hilb3,Hilb4}. The operator $\mathcal{R_\rho}$ is not a usual operator, but its trace values behave similarly to what was expected by the ETH.
Although $\mathcal{R}$ is not a density matrix, it is possible to calculate its Wigner transform by expanding it in the Fock basis, 
\begin{equation}
    \mathcal{R}_\rho (t)=\sum_{n,m}r_{n,m}(t)\ket{n}\bra{m};
    \label{R}
\end{equation}
and its Wigner transformation is

\begin{equation}
    W\{\mathcal{R}_\rho\}(t)=\sum_{n,m} r_{n,m}(t) \Pi_{n,m},
    \label{WR}
\end{equation}    

where $\Pi_{n,m}=W(\ket{n}\bra{m})$. Figure \ref{WIG2} shows the surface plot of the Wigner transform of $\mathcal{R}_\rho$, as well as their components separated in diagonal contribution and non-diagonal contribution, for $t=5\pi$ and $ t=10 \pi$. The equivalent contour plot of \ref{WIG2} is shown in figure \ref{WIG3}. As it is clear from these figures, the Wigner transform of  $\mathcal{R}_\rho$ is dominated by the non-diagonal terms. The non-positivity of the diagonal terms of the Wigner transform can be interpreted as non-classical signatures, and they are responsible to the fluctuations of the overlap. On the other hand, the diagonal terms are not sensible to the complex structure in phase space that is relevant also in sub-Planck scale. 

\section{Conclusion}

The death of the overlap , a universal behavior of quantum chaotic systems, is also extensive to non-periodic dynamics, as has been shown. The time variance of overlap has shown to be useful to distinguish the dynamics from periodic to non-periodic, which is usually a hard task. It became evident the overlap collapse does not necessarily imply that the system is classical, but it can be regarded as an effective classical behavior, i.e. although the state is strongly non-classical. The classicality is due to the observable choice, but it does not  mean that the classical regime has been achieved. By analyzing the Wigner function, it was shown that the dynamics is strongly affected by the non-diagonal of the density matrix, and it affects the overlap. The overlap operator dynamics has a complex structure, i.e., its Wigner transform has significant changes in scales below $\hbar/2$. Although the diagonal part of the overlap operator may exhibit negative values in its Wigner transform, much of the information about its surface is in the non-diagonal part. Consequently, while the overlap is a good indicator of aspects of the periodicity of the expected value dynamics of the position and momentum operators, it is not a good representative of the dynamics as a whole. As the overlap depends only on the diagonal terms of the overlap operator, then it is not sensible to the sub-plank structure.


	\section*{Acknowledgments}
	The authors gratefully acknowledge the support of the Brazilian Agency Funda\c{c}\~{a}o de Amparo \`{a} Pesquisa do Estado de Minas Gerais
	(FAPEMIG), the support provided by the Brazilian Agency Coordena\c{c}\~{a}o de Aperfei\c{c}oamento de Pessoal de N\'{i}vel Superior (CAPES) and the support of the National Council for Scientific and Technological Development – CNPq.


\begin{thebibliography}{99}

\bibitem{Bal} L. E. Ballentine, Y. Yang, and J. P. Zibin, Phys. Rev. A 50,
(1994) 2854.

\bibitem{Ball98} L.~E. Ballentine and S.~M. McRae, \newblock Phys. Rev. A
\textbf{58}, 1799 (1998).

\bibitem{Ball01} L.~E. Ballentine, \newblock Phys. Rev. A \textbf{63}, 024101
(2001).


\bibitem{Ball2005} N. Wiebe and L. E. Ballentine, Phys. Rev. A \textbf{72},
022109 (2005).

\bibitem{Kirchmair} G. Kirchmair, et al., Nature 495 (2013) 205.

\bibitem{Bosco2016} A.R.Boscode Magalhães, A. C. Oliveira, Physics Letters A
380 (2016) 554-561.



\bibitem{Oliveira2021} Oliveira, A. C.,
Continuum reset dynamics as a pathway to Newtonian classical limit of Quantum Mechanics,
Physica A,Volume 579, (2021), 126099.


\bibitem{Oliveira2014} A. C. Oliveira, Physica A 393 (2014) 655--668.
\bibitem{Gampel} Gampel, F. and Gajda, M., Phys. Rev. A 107, 012420, (2023).

\bibitem{adelcioJMP} A. C. Oliveira, Journal of Modern Physics, \textbf{3,}
694 (2012).
\bibitem{Oliveira09b} Oliveira, A. C. and de Magalhaes, A.R.Bosco, %
\newblock Phys. Rev. E, \textbf{80}, 026204, (2009).
\bibitem{Angelo} R. M. Angelo, Phys. Rev. A 76, 052111 (2007).
\bibitem{Zur1} Zurek, W. H., S. Habib, and J. P. Paz. 1993. Coherent States via Decoherence. Phys. Rev. Lett. 70 (9): 1187

\bibitem{adelcio2012} A. C. Oliveira, and A.R. Bosco de Magalhães, and
J. G. Peixoto Faria, Physica A, \textbf{391, }5082 (2012).

\bibitem{Ball3} L. E. Ballentine and S. M. McRae. Moment equations for probability distributions in classical and quantum mechanics. Phys. Rev. A, 58(3):1799, 1998.

\bibitem{Ball4} L. E. Ballentine. Lyapunov exponents for the differences between quantum and classical dynamics. Phys. Rev. A, 63(2):024101, 2001.

\bibitem{Ball5} Nathan Wiebe and L. E. Ballentine. Quantum mechanics of Hyperion. Phys. Rev. A, 72(2):022109, 2005.


\bibitem{Ball2} Leslie E. Ballentine, Yumin Yang, and J. P. Zibin,
Inadequacy of Ehrenfest’s theorem to characterize the classical regime.Phys. Rev. A, \textbf{50(4)} 2854 (1994)






\bibitem{Echo1} Peres, Asher, Stability of quantum motion in chaotic and regular systems, Phys. Rev. A, Volume 30, (1984), 1610.


\bibitem{Echo2} Diego A. Wisniacki Phys. Rev. E 67, (2003), 016205.

\bibitem{Echo3} Arseni Goussev et al. (2012) Loschmidt echo. Scholarpedia, 7(8):11687.

\bibitem{Rev1} T. Gorin, T. Prosen, T. H. Seligman, and M. Znidaric, Phys. Rep. 435, 33 (2006).

\bibitem{Rev2}  P. Jacquod and C. Petitjean, Adv. Phys. 58, 67 (2009).

\bibitem{Prosen} Tomaz Prosen and Marko Znidaric. Stability of quantum motion and correlation decay, 	J.Phys.A 35, 1455 (2002).


\bibitem{Fanizza} M. Fanizza et al. ,Phys. Rev. Lett. 124, (2020) 060503.

\bibitem{Vedral} V. Vedral, Rev. Mod. Phys. 74,(2002) 197.

\bibitem{Bartlett} S. D. Bartlett, T. Rudolph, and R. W. Spekkens, Rev. Mod. Phys. 79, 555 (2007).


\bibitem{Herrada2023} Zarate-Herrada DA, Santos LF, Torres-Herrera EJ. Generalized Survival Probability. Entropy (Basel). 2023 Jan 20;25(2):205. doi: 10.3390/e25020205. PMID: 36832572; PMCID: PMC9955597.


\bibitem{Ber77} G.~P. Berman and G.~M. Zaslavsky, \newblock Physica A
(Amsterdam) \textbf{91}, 450 (1977).

\bibitem{Ber81} G.~P. Berman, A.~M. Iomin, and G.~M. Zaslavsky, \newblock %
Physica D, \textbf{4}, 113 (1981).

\bibitem{Oliveira06} A.~C. Oliveira and J. G. Peixoto de Faria and M.~C.
Nemes, \newblock Phys Rev. E \textbf{73}, 046207 (2006).


\bibitem{Adelcio2003} Oliveira,A.C., Nemes,M.C. and Romero, K.M.Fonseca,
\newblock Phys Rev. E \textbf{68}, 036214 (2003).

\bibitem{Faria} J. G. Peixoto de Faria, \newblock The European Physical
Journal D, \textbf{42}, 153, (2007).


\bibitem{Ber91} G.~P. Berman and V. Yu Rubaev and G.~M. Zaslavsky, \newblock %
Nonlinearity \textbf{4}, 543 (1991).

\bibitem{Leonski1996} W.~Leo{\'n}ski, Quantum and classical dynamics for a
pulsed nonlinear oscillator, Physica A 233~(1-2) (1996) 365--378.

\bibitem{Iomin2001} A. Iomin and G. M. Zaslavsky, \newblock Phys Rev. E
\textbf{63},047203 (2001).

\bibitem{Iomin2003} A. Iomin and G. M. Zaslavsky, \newblock Phys Rev. E
\textbf{67},027203 (2003).







\bibitem{Hilb1} R. A. Jalabert and H. M. Pastawski. “Environment-Independent Decoherence Rate in Classically Chaotic Systems”. Phys. Rev. Lett., Vol. 86, pp. 2490, 2001.

[ Jacq 01, Cerr 02, Wisn 02, Cerr 03, Wisn 03, Cucc 04, Gori 04, Guti 09]

\bibitem{Hilb2} Ph. Jacquod, P. G. Silvestrov, and C. W. J. Beenakker. “Golden rule decay versus Lyapunov decay of the quantum Loschmidt echo”. Phys. Rev. E, Vol. 64, p. 055203(R), 2001.

\bibitem{Hilb3} D. A. Wisniacki and D. Cohen. “Quantum irreversibility, perturbation independent decay, and the parametric theory of the local density of states”. Phys. Rev. E, Vol. 66, p. 046209, 2002.

\bibitem{Hilb4} M. Guti´errez and A. Goussev. “Long-time saturation of the Loschmidt echo in quantum chaotic billiards”. Phys. Rev. E, Vol. 79, p. 046211, 2009.

\bibitem{Pegg}  D. T. Pegg and S. M. Barnett, J. Mod. Opt. 44, 225 (1997).


\bibitem{NCMRev} Oliveira, A. C., Nonclassicality measures of single-mode quantum states, International Journal of Quantum Information  2025 23:04, 2530002.


\bibitem{Dirac} P. A. M. Dirac, R. H. Fowler, The quantum theory of
dispersion, Proc. R. Soc. Lond. 114, 710 (1927).

\bibitem{Lemos2018} Humberto C.F.Lemos, Alexandre C.L.Almeida, Barbara
Amaral, Adelcio C.Oliveira, Physics Letters A , 382 (2018), 823-836.

\bibitem{Karen2} K. M. Fonseca-Romero, M. Reis, A. C. Oliveira, Physics Letters A, V. 486,
129097 (2023).







\bibitem{Reis2018} M. Reis e Silva Jr and A. C. Oliveira, SBFoton International Optics and Photonics Conference (SBFoton IOPC),
(2018) 1-5.

\bibitem{Reis2021} Mauricio Reis, Adélcio C. Oliveira, Complementary Resource Relation of Concurrence and Roughness for a two Qubits State, 	arXiv:2106.00036.


 \bibitem{Reis2021b} Reis, M., Oliveira, A.C. Relation between the roughness, linear entropy and visibility of a quantum state, the Jaynes–Cummings model. J Comput Electron 20, 2189–2198 (2021). https://doi.org/10.1007/s10825-021-01761-0



\bibitem{ReisQutip} M. Reis, Numerical implementation of roughness, a function to measure the quantumness of a quantum state. https://github.com/mreis-phys/quantum-roughness


\bibitem{Thermal6} Ueda, M. Quantum equilibration, thermalization and prethermalization in ultracold atoms. Nat Rev Phys 2, 669–681 (2020). https://doi.org/10.1038/s42254-020-0237-x





















































\end{thebibliography}
\end{document}